\documentclass[12pt,graphics] {article}
\usepackage[dvips]{graphicx,color}
\newcommand{\bea}{\begin{eqnarray}}
\newcommand{\eea}{\end{eqnarray}}
\newcommand{\be}{\begin{equation}}
\newcommand{\ee}{\end{equation}}

\def\llb{{\Bigg {\lbrack}}\!\!{\Bigg {\lbrack}}}
\def\rrb{{\Bigg {\rbrack}}\!\!{\Bigg {\rbrack}}}
\def\lsim{\mathrel{\lower2.5pt\vbox{\lineskip=0pt\baselineskip=0pt
          \hbox{$<$}\hbox{$\sim$}}}}
\def\gsim{\mathrel{\lower2.5pt\vbox{\lineskip=0pt\baselineskip=0pt
          \hbox{$>$}\hbox{$\sim$}}}}

\begin{document}

\begin{titlepage}
\begin{trivlist}\sffamily\mdseries\large
\item
MTR 01W0000017\\[-0.8ex]
\hrule ~\\[-1.2ex]
{\mdseries MITRE TECHNICAL REPORT}\\[1cm]
\LARGE
\begin{center}
\bfseries
Secrecy, Computational Loads and Rates in\\
Practical Quantum Cryptography\\[2.5cm]
\end{center}
\mdseries
\large
G. Gilbert\\[0.8ex]
M. Hamrick\\[0.4ex]
~\\
\large
\textbf{May 2001}\\[3.1cm]
\begingroup\footnotesize
\begin{tabbing}
\textbf{Sponsor:} \phantom{spo} \= The MITRE Corporation \phantom{phantomphantomprospo} \=
\textbf{Contract No.:} \phantom{pro}\= DAAB07-01-C-C201 \\
\textbf{Dept. No.:} \>W072 \>\textbf{Project No.:} \>51MSR837\\[0.6cm]
The views, opinions and/or findings contained in this report \> \>
   Approved for public release; \\
are those of The MITRE Corporation and should not be \> \>
   distribution unlimited.  \\
construed as an official Government position, policy, or \\
decision, unless designated by other documentation. \\[0.3cm]
\copyright 2001 The MITRE Corporation
\end{tabbing}
\endgroup
~\\
\bfseries
\Large
MITRE\\
\normalsize
Washington ${\mathbf C^3}$ Center\\
McLean, Virginia\\
\clearpage
\end{trivlist}
\end{titlepage}

\Large
\begin{center}

{\bf Secrecy, Computational Loads and Rates in Practical Quantum Cryptography$^{\ast}$}
\normalsize

\vspace*{45pt}
Gerald Gilbert$^{\dag}$ and Michael Hamrick$^{\ddag}$\\
The MITRE Corporation\\ McLean, Virginia 22102
\end{center}

\begin{abstract}

A number of questions associated with practical implementations of
quantum cryptography systems having to do with unconditional
secrecy, computational loads and effective secrecy rates in the
presence of perfect and imperfect sources are discussed. The
different types of unconditional secrecy, and their relationship
to general communications security, are discussed in the context
of quantum cryptography. In order to actually carry out a quantum
cryptography protocol it is necessary that sufficient
computational resources be available to perform the various
processing steps, such as sifting, error correction, privacy
amplification and authentication. We display the full computer
machine instruction requirements needed to support a practical
quantum cryptography implementation. We carry out a numerical
comparison of system performance characteristics for
implementations that make use of either weak coherent sources of
light or perfect single photon sources, for eavesdroppers making
individual attacks on the quantum channel characterized by
different levels of technological capability. We find that, while
in some circumstances it is best to employ perfect single photon
sources, in other situations it is preferable to utilize weak
coherent sources. In either case the secrecy level of the final
shared cipher is identical, with the relevant distinguishing
figure-of-merit being the effective throughput rate.

\end{abstract}

\normalsize
~\\[1.75in]
\hrule width 2.5 in
\vspace*{.175in}
{\footnotesize
$\ast$ This research was supported by MITRE under MITRE Sponsored
Research Grant
51MSR837.\\
\dag ~ggilbert@mitre.org\\
\ddag ~mhamrick@mitre.org
}
\clearpage

\section{Introduction}

Recent progress in telecommunications and optoelectronics is
making it possible for serious consideration to be devoted to the
prospect of practical quantum key distribution systems. One
possible use to which quantum key distribution may be put is as a
precursor to the real-time encryption of plaintext using the
method of the Vernam cipher, or one-time pad. Such an end-to-end
secret system, which would offer the promise of undecipherable
communications, is attractive for a wide class of applications in
spite of the fact that Vernam encryption requires a cipher that is
bit-for-bit as long as the plaintext. However, in order to make
progress in implementing real systems that will work in actual
physical environments, it is necessary to determine to what extent
predictions regarding secrecy in ideal situations can be realized
with practical devices and components. To that end, in this paper
we clarify the important but often overlooked distinction between
secrecy and security in communications. This is a distinction {\it
with} a difference, as it illuminates the proper constraints that
must be satisfied in order to achieve privacy in realistic
communications scenarios. We then focus on the specific types of
secrecy that are relevant to quantum cryptogaphy, distinguishing
between what may actually be achieved in practical implementations
in physical environments and what is restricted to formal
discussions that are pertinent only to idealized models. We point
out that there are actually three ``levels" of unconditional
secrecy that one may discuss. We provide a careful description of
a particular implementation of sifting, error correction and
privacy amplification for three reasons: it is of intrinsic
interest, it identifies the points at which authentication is
required to prevent man-in-the-middle attacks by an eavesdropper,
and it provides the necessary input for the detailed result we
subsequently present on computational loads. These are the
requirements on classical computing machinery that must be
satisfied in order to actually carry out the quantum cryptographic
transmission. Finally, we consider practical implementations of
quantum cryptography in which optical fiber is used as the quantum
channel and compare the use of perfect single photon sources with
the use of weak coherent sources, and we in particular consider
this in distinct scenarios in which the eavesdropping attacks are
constrained to lesser and greater degrees. Based on this we are
able to identify situations in which it is preferable to use a
perfect single photon source instead of a weak coherent source,
and {\it vice versa}.

\section{Quantum Key Distribution Using the BB84 Protocol}

Here we provide a very brief description of the basic elements of quantum
key distribution. We will illustrate this with the original four-state QKD protocol
developed by Bennett and Brassard in 1984 known as the ``BB84" protocol \cite{BB84}.
For definiteness in this illustration
we will assume that individual photons serve as the quantum bits for the protocol,
or more precisely, the polarization states of individual photons. To carry out the
protocol one of the parties transmits
a sequence of photons to the other party.
The parties publicly agree to make use of two distinct polarization bases which are
chosen to be maximally non-orthogonal. In a completely
random order, a sequence of photons are prepared in states of definite polarization
in one or the other of the two chosen bases and transmitted by one of the parties to the
other through a channel that preserves the polarization. The photons are measured by
the receiver in one or the other of the agreed upon bases, again chosen in a
completely random order. The choices of basis made by the transmitter and receiver
thus comprise two independent random sequences.
Since they are independent random sequences of binary numbers, about half of the basis
choices will be the same and are called the ``compatible" bases, and the other half
will be different and are called the ``incompatible" bases.
The two parties compare publicly, making use for this purpose of a classical
communications
channel, the two independent random sets of polarization {\it bases} that were used,
without revealing the polarization {\it states} that they observed.
The bit values of those polarization states measured in the compatible bases
furnish the ``sifted key."
Note that, if the two parties used classical signals
to send the key, an eavesdropper
could simply measure the signals to obtain complete knowledge of the key. If, on the
other hand, the two parties use
single photons to transmit the key, the Heisenberg Indeterminacy Principle guarantees
that an eavesdropper cannot measure the polarizations
without being detected.
The sifted keys possessed by each of the parties will in general
be slightly different from each other due to errors caused by the use of imperfect
equipment.
A classical error correction procedure, carried out through the classical communication
channel, is executed in order to produce identical, error-free keys at both ends.
It is possible that an enemy may have obtained some information about the key during
the publicly-discussed error correction phase of the protocol.
In addition, it is also possible for the enemy to have obtained information due to the
presence in the sequence of quantum bits of multiple photon states.
The process of ``privacy amplification" is therefore applied to the sifted, error-free
key.  This has the effect of reducing the information available to the enemy to an
arbitrarily low value with extremely high probability.

\section{Security and Secrecy}
\label{S:ss}

Much of the current research in quantum cryptography is devoted to the construction of
proofs of the ``unconditional security" of quantum cryptographic protocols.
The first such proofs assumed idealized equipment
\cite{bb84proofs3, bb84proofs4, bb84proofs2, bb84proofs1}.  More
recent investigations have explored the consequences of the noise, errors, and losses
that inevitably occur in a practical implementation
\cite{lutkenhaus1, lutkenhaus-T, gh}.
Our concern here is to
understand what, precisely, is meant by ``unconditional security", to develop a
nomenclature that is more consistent with established definitions, and to point out
the fundamental way in which quantum cryptographic security differs from what is
achievable with classical cryptographic techniques.

\subsection{The Distinction between Communications Security and Cryptographic Secrecy}
\label{nsa}

First, we note that the distinction between ``security" and ``secrecy" is
frequently glossed over.  Shannon's original
definition of the term ``perfect secrecy" is found in his
seminal work on the subject:
{\it Communication Theory of Secrecy Systems} ({\it cf} \cite{Shannon}).
The basic requirement for secrecy is that, in comparing the situation {\it before}
the enemy has intercepted the transmission with the situation
{\it after} any such interception (and analysis) has occurred,
the {\it a posteriori} and {\it a priori}
probabilities for the enemy to know the content of the transmission must
be identical.  Under this definition, the transmission of a key $K$ from the
transmitter (Alice) to the receiver (Bob) is
perfectly secret when

\be
I(K, X_1) = I(K, X_2)~,
\ee

where $I$ denotes the mutual information, $X_1$ is the data the eavesdropper (Eve) has
obtained prior to the execution
of the key distribution protocol
and $X_2$ is the data she has after the protocol.  We see that the term
 ``secrecy" applies solely to
the protection provided by the cryptographic protocol alone.  The system taken as a whole
may not be {\it secure} even though the protocol is perfectly {\it secret}. This may
happen in
one of two ways.  Either Eve has prior knowledge of the key, so that $I(K, X_1) > 0$, or
Eve manages to add to her information the key at some later time, thus obtaining a
string $X_3$ with $I(K, X_3) > 0$.  In either case, Eve has obtained information without
compromising the protocol itself.

The notion of ``security" or, more precisely, communications
security~\cite{fs1037}, is defined in a broader context to which
Shannon's ``perfect secrecy" is a contributing factor. According
to the standard scheme advocated by the U.S. National Security
Agency in \cite{nsadoc}, communications security is split into
four separate categories:

(1) cryptosecurity - [The] component of communications security that results from the
provision of {\it technically sound cryptosystems}
(emphasis added) and their proper use.

(2) emission security - Protection resulting from all measures taken to deny unauthorized
persons information of value which might be derived from intercept and analysis of
compromising emanations from crypto-equipment, computer and telecommunications systems.

(3) physical security - [The] component of communications security
that results from all physical measures necessary to safeguard
classified equipment, material, and documents from access thereto
or observation thereof by unauthorized persons.

(4) transmission security - [The] component of communications security that results from
the application of measures designed to protect transmissions from interception and
exploitation by means other than cryptanalysis.

In this paper we will use the word {\it secrecy} as synonymous with
cryptosecurity, in the sense of definition (1) above, in recognition of the fact
that the security of the entire cryptographic system depends on several factors in
addition.

{\it Unconditional secrecy} refers to secrecy that remains intact when the
cryptosystem is subjected to attacks by an enemy equipped with
unlimited time and - within the
constraints dictated by the laws of physics - unlimited computing machinery.  We
therefore assume that Eve has complete access to the specifications of the protocols
used by Alice and Bob as well as physical access to both the quantum and classical
channels.  Secrecy is assured by fundamental restrictions of quantum mechanics.
Note that unconditional secrecy is completely beyond the reach of classical
cryptographic systems. Although Vernam encryption
is of course unbreakable without explicit knowledge of the key,
there is no classical, purely {\it cryptographic} method of distributing the key
which provides perfect unconditional secrecy (the use of a trustworthy
courier, for instance, does not correspond to cryptographic secrecy, but rather is
an example of transmission security).
With regard to classical cryptographic protection applied to key distribution,
if Eve has free access to
the classical communications channel and unlimited computing power, there is no
way to guarantee an upper bound to the information she is able to obtain.  Such bounds
can be obtained for quantum cryptographic systems that rely on the distribution of
Vernam keys over a quantum channel.  Information bounds resulting from proofs of
secrecy fall into three categories.  We will call these
perfect secrecy,
asymptotic secrecy, and secrecy in the sense of privacy
amplification.

\subsection{Three Categories of Secrecy}

In this section we describe the secrecy of quantum key distribution in terms
of the mutual information $I(K,X)$ of the key $K$ and a string containing the
information Eve has obtained $X$. Note that the uniformity of the key, that is, the
uniformity of the probability distribution of the
key in the
space of all possible keys, is also a requirement
for successfully using the key in Vernam encryption.  In practice, both
secrecy and uniformity can be established by bounding Eve's entropy:

\be
H(K|X) \geq {\cal L}\left( K \right) - \epsilon
\ee

where ${\cal L}\left( K \right)$ is the length of the key string in bits.
Given this bound on the entropy, we have

\be
I(K,X) \equiv H(K) - H(K|X) \leq {\cal L}\left( K \right) - H(K|X) \leq \epsilon
\ee

which establishes the secrecy bound and

\be
H(K) = I(K,X) + H(K|X) \geq H(K|X) \geq {\cal L}\left( K \right) - \epsilon
\ee

which establishes uniformity of the key in the space of binary strings of length
${\cal L}\left( K \right)$.  In this paper we will restrict our attention to
the secrecy of
the key, with the implicit understanding that the corresponding statements of
uniformity also apply.

\subsubsection{Perfect secrecy}

Proofs of {\it perfect secrecy} often proceed by assuming that
Eve has no initial
knowledge of the key, so that $I(K, X_1) = 0$, and then attempting to show
that $I(K, X_2) = 0$.
It should be noted that, in view of the distinction between secrecy and security as
clarified in Section \ref{nsa} above, such an approach, although common, is not
completely general. Instead, one should proceed by not assuming anything about the
{\it specific} amount of information that Eve may have previously obtained. Thus, a proof
of perfect secrecy should begin merely with the assumption that Eve has {\it some}
amount of {\it a priori}
knowledge of the key, say an amount $\hat I$ which may be zero or greater than zero,
so that $I(K, X_1) = \hat I$.
(Of course, if $\hat I>0$ then there has been some non-cryptographic violation of the
security of the communication system, arising from failure to observe proper ``technically
sound cryptosystem design and practice," as discussed in \cite{gh}. The communications
system in this case is already insecure, even though the cryptographic protocol that it
employs may be provably secret.)
A proof of
perfect secrecy then consists in showing that $I(K, X_2) = \hat I$ as well, so that

\be
I(K, X_2)-I(K, X_1)\equiv\Delta I=0~.
\ee

{\it Unconditional} perfect secrecy consists
in further demonstrating that $\Delta I=0$ in the absence of any assumptions about the
computational resources and capabilities of Eve. This more general approach
includes as a special case those proofs which rely on the restrictive initial
assumption that
$I(K, X_1) = 0$. The point here is that one may indeed demonstrate that a
{\it cryptographic protocol} (in this case, key distribution) is
perfectly secret, {\it i.e.} $\Delta I=0$, and even show that it is unconditionally
perfectly secret, {\it i.e.} $\Delta I=0$ with no assumed conditions, and at the same time
have the actual {\it communications} be entirely insecure.
However, all this is moot since, in any event, perfect unconditional
secrecy ({\it i.e.}, having $\Delta I=0$ hold {\it exactly}) is not
achievable for {\it any} cryptographic protocol
in which strings of finite length are transmitted, even for perfect quantum
systems ({\it i.e.}, systems utilizing perfect singlet state quantum bits),
since there is a finite non-vanishing
probability that Eve can eavesdrop on the quantum channel without introducing
errors, thus gaining information without being detected.

\subsubsection{Asymptotic secrecy}

In practice, the mutual information is not strictly zero, but is bounded by
a quantity that is exponentially small.  For the case of {\it asymptotic
secrecy\/}, the bound is of the form (for brevity
here and in Section \ref{S:uspa} we now {\it do} assume that $I(K, X_1) = 0$)

\be
\label{E:asyminfo}
I(K, X_2) \leq 2^{- {\cal O} \left( N_s \right) }~,
\ee

where the exponent
is of the order of the size in bits of the resulting key, $N_s$, and the inequality holds
with probability close to 1.

For practical purposes there is little difference between asymptotic
and perfect secrecy, since the exponential
quantity that bounds $I(K, X_2)$ is extremely small.
Note that in the limit of an infinitely long cipher we
have $\lim_{N_s\rightarrow\infty}2^{-{\cal O} \left( N_s \right)}=0$
and we recover perfect secrecy as defined above
(here in the special case $I(K, X_1) = 0$).

\subsubsection{Secrecy in the sense of privacy amplification}
\label{S:uspa}

Discussions of the secrecy of practical implementations of quantum cryptography
\cite{lutkenhaus1, lutkenhaus-T, gh} typically establish {\it
secrecy in the sense of privacy amplification\/}.
This is because information leaked to Eve
during error correction and other phases of the protocol is removed by a
privacy amplification protocol \cite{bbcm}.  This results in a bound of the form

\be
\label{uspa}
\langle I(K, X_2) \rangle < 2^{-g_{pa}}~,
\ee

where the angle brackets indicate an expectation value over the class of hash functions
used to carry out privacy amplification, and where $g_{pa}$ is a security
parameter determined by the protocol, independent of the size of the
string, $N_s$.  The price we pay for this is a shortening of the string by
a number of bits equal to the upper bound on Eve's mutual information prior to
privacy amplification plus the security parameter $g_{pa}$.

Privacy amplification was first applied to the problem of quantum cryptography
by Bennett {\it et al.\/} \cite{BB84}.  Closely allied techniques are used in
security proofs by Mayers \cite{bb84proofs3} and by
Biham {\it et al.\/} \cite{bb84proofs2} which treat quantum key distribution systems
using idealized single photon sources, but which allow Eve to make any possible quantum
mechanical attack.   Gilbert and Hamrick \cite{gh} apply privacy amplification to
practical systems using realistic photon sources subject to ``individual" attacks, in
which Eve attacks each photon independently of the others.

Secrecy in the sense of privacy amplification is different from asymptotic secrecy in
three important respects.
First, there is no necessary relationship between the security parameter $g_{pa}$ and the
size of the key, $N_s$.  Second, and corollary to the first difference, secrecy in the
sense of
privacy amplification does not necessarily become perfect secrecy in the limit of large
keys.   (This would happen if $g_{pa}$ were arbitrarily chosen to be some fraction of the
key size.  Since $g_{pa}$ bits are removed from
the string by
privacy amplification, this approach is costly in terms of the rate of generation of key
material.  Note that bounds of this form are in fact
obtained by Biham {\it et al.\/} \cite{bb84proofs2}.)
Third, we improve the secrecy bound in privacy amplification by
increasing $g_{pa}$, thus producing a {\it shorter} key.
In contrast, we improve the secrecy bound of an asymptotically secret
string simply by constructing a {\it longer} key.

Finally, note that the asymptotic secrecy bound is expressed as an
absolute bound on the mutual information, while the secrecy bound for privacy
amplification is given as an average over the set of hash functions.
This is really an artificial distinction, since absolute bounds can be derived
for the privacy amplification result as well \cite{lutkenhaus-practical, ght}.

There is no qualitative
distinction between the secrecy achieved by systems using
pure single photon sources and that achieved by systems whose sources generate a mixture
of single-photon and multi-photon
pulses.  In both cases information will be leaked to Eve at some point in the protocol,
and any protocol that removes the leaked information by
classical privacy amplification falls into the category of
secrecy in the sense of privacy amplification.  We shall return to this point
later.

It must be emphasized that the distinction between security and secrecy is not
merely a matter of semantics, but has a definite impact on what is required in order
to obtain proofs of unconditional secrecy.  In particular,
unconditional proofs of secrecy need not
be concerned with eavesdropping attacks that compromise the physical security of the
cryptographic system.  As an example, attacks in which Eve manipulates the efficiency
of Bob's detection device have been discussed in the literature \cite{brassard-eta}.
This manipulation might be achieved by modifying the wavelength of the pulses received
by Bob, in
which case the attack can be countered by inserting a narrowband filter in the optical
path.  Otherwise the attack would require direct access to Bob's detection devices, which
is a breach not of cryptosecurity but of physical security.  Note also
that, given such access to Bob's physical installation, Eve could compromise
the security of the entire system much more easily by using other parts of Bob's equipment,
notably the mass storage where the secret key material is kept prior
to use.  Such attacks are clearly beyond the scope of what is
required for proofs of unconditional secrecy.

\section{Error Correction, Authentication, Privacy Amplification and Computational Loads}

The BB84 protocol is really of family of protocols that can be implemented in many
different ways.  The details of the quantum transmissions depend on the way Alice and
Bob choose to represent qubits, {\it e.g.} by choosing to use photons or electrons, and by
the quantum basis states they pick to encode the information.  Similarly the classical
transmissions used to identify basis choices, correct errors, and authenticate the
communications channels can be implemented in a variety of ways.  This section describes
details of a specific implementation to give an idea of the kind of classical
communications and computational algorithms that are required.  See reference
\cite{gh} for
a more detailed analysis of the computational and communications resources required to
carry out this  implementation of BB84.

The complete implementation of the protocol occurs in three phases.
The first phase is the production
of a sifted string of bits shared, except for some errors, by Alice and Bob. This is
achieved by applying the BB84 protocol previously described. If the equipment
were perfect and there were no possibility of errors, the sifted strings would be identical
and any errors would be due to an attempt by Eve to measure the polarizations of the
photons.  Since real equipment is never perfect,
it is essential to include mechanisms to correct the errors due to the
equipment and to
eliminate information leaked to Eve in the process. The second phase is thus error
correction. Alice and Bob agree on a systematic protocol to identify and correct the
errors. Since Eve can eavesdrop on this discussion, some additional amount of key
information is leaked to her at this time.
The third phase is privacy amplification, during which Alice and Bob apply a hash
transformation to the error-corrected string.  This results in a shorter string about
which Eve's expected information is vanishingly small. At various points during
these three phases, Alice and Bob must authenticate their communications to ensure
that Eve is not making a man-in-the-middle attack.

In addition, Alice and Bob check that the observed error rate is below a
threshold value.  The information that Eve can obtain by directly measuring
single photon pulses is bounded by the error rate on the channel, and the protocol uses
privacy amplification to protect against information losses up to that bound.
By testing the error rate, Alice and Bob can detect any attempt by Eve to obtain
additional information by a stronger direct-measurement attack on the quantum channel.

\subsection{Sifting}

The first phase of the key distribution protocol is the generation of
an initial sifted string that is shared between Alice and Bob, but which may
contain errors and about which Eve may have partial information.  We describe here a
specific implementation of the sifting protocol.
Alice generates two blocks of $m$ random bits.  The first block is the
raw key material, and the second block determines the choice of basis she uses
to transmit the bits over the quantum channel. Bob generates a single block of
$m$ bits that reflect his choice of basis in measuring the incoming qubits. Bob
must now identify to Alice those pulses for which he detected a qubit and inform
her of his choice of basis for those pulses.  Bob has several choices available in
deciding how he wants to encode this information.  For purposes of estimating the
computational load it is necessary to choose a
specific implementation.  Accordingly, we
choose an implementation in which Bob sends to Alice two
pieces of information for each photon he detects.  The first piece indicates which of the
$m$ bits sent by Alice resulted in the detected photon, and the second gives Bob's choice
of basis for that photon.
This requires that Bob send $2n\left(1+\log_2m\right)$ bits for each block of key
material.  (The factor of 2 arises from the fact that Bob measures
about one half of the
qubits in the wrong basis.  These qubits are discarded to
produce the sifted string.  Since the sifted string is of length $n$, the
number of qubits detected by Bob is $2n$.)

Once Alice has received Bob's information, she compares Bob's basis choices
with her own and informs Bob of the results.  Alice can accomplish this by
sending Bob a single bit corresponding to each of the photons Bob detected,
resulting in a total of $2n$ bits of information sent to Bob.

\subsection{The Need for Authentication during Sifting}

We must now augment the protocol with provisions that will prevent
Eve from making the so called man-in-the-middle attack.  In this
attack, Eve interposes herself between Alice and Bob, measuring
Alice's pulses on the quantum channel as though she were Bob, and
generates a distinct set of pulses to send to Bob as though she
were Alice.  In all her subsequent correspondence with Alice over
the classical channel, she responds just as Bob would, and in all
correspondence with Bob she plays the role of Alice.  After the
first phase of the protocol, Eve has two blocks of sifted keys,
one of which she shares with Alice and the other with Bob.
Assuming she can continue this attack through the error correction
and privacy amplification phases, she will have completely
compromised Alice and Bob's ability to use the keys to transmit
secret information.  At this point Eve is able to decipher any
encrypted information sent between Alice and Bob, always passing
the re-encrypted text to the intended recipient so that neither
Alice nor Bob is any the wiser.

In order to prevent this state of affairs, it is necessary to
provide an authentication mechanism to guarantee that the
transmissions received by Bob were sent by Alice, and not by Eve,
and to guarantee that the transmissions received by Alice were
sent by Bob. Wegman and Carter \cite{wc} describe an
authentication technique based on ``$almost~universal_2$" sets of
hash functions that are well suited to this problem.  The
authentication works as follows. Alice and Bob first agree upon a
suitable space of hash functions to be used for authentication.
All details of their agreement may be revealed to Eve without
compromising the authentication.  For each message that is to be
authenticated, Alice picks a hash function from the space that is
known to Bob, but not to Eve.  She does this by using a string of
secret bits that is known only to herself and Bob as an index to
select the hash function. She uses some of the secret key
generated by previous iterations of the protocol to provide this
secret index, with the result that some of the key material is
sacrificed in order to achieve authentication.  She then applies
the hash function to the block of raw data to produce an
authentication key. This authentication key is transmitted to Bob
along with the message.  Bob uses the same string of secret bits
to pick the same hash function, applies it to the message, and
compares the result with the authentication key sent by Alice. If
they match, Bob concludes that Alice, and not Eve was, the sender
of the message. Wegman and Carter describe a class of hash
functions such that the probability that Eve can generate the
correct authentication key without knowing the index used is
vanishingly small. Let ${\cal M}_1$ denote the precondition that
Eve has obtained a copy of the message to be authenticated and
${\cal T}^{(E)}_1$ denote the outcomes in which Eve guesses the
tag for the message. The probability of such an outcome is

\be
\label{E:authprob}
{\cal P}({\cal T}^{(E)}_1|{\cal M}_1) = 2^{-g_{auth}}~,
\ee

where $g_{auth}$ depends on the space of hash functions Alice and
Bob have chosen to use for the protocol.  It can be made as large as desired
by making the space sufficiently large.  Alice and Bob do pay a price for increased
confidence. A larger space of functions requires a larger set of indices, and thus a
longer string of secret bits must be sacrificed to perform the authentication.  The
other restriction on the protocol is that a new hash function, and thus a new index,
must be used for each message to be authenticated if we desire to maintain this
upper bound on Eve's ability to spoof the authentication process. If we allow Eve
to obtain one prior message and tag, denoted as ${\cal M}_1{\cal T}_1$, and then
allow her to obtain the next message, denoted as ${\cal M}_2$, as well as the
information that Alice and Bob intend to use the same hash function
for both, her chances of guessing the second tag improve only slightly to

\be
{\cal P}({\cal T}^{(E)}_2|{\cal M}_1{\cal T}_1{\cal M}_2) = 2^{1-g_{auth}}~.
\ee

If we allow additional messages to be authenticated using the same
hash function, Wegman and Carter's analysis provides no upper bound on Eve's
ability to produce a correct authentication tag.  Although it would be more
efficient to allow the same hash function to be applied exactly twice, we will
consider the simpler case in which a new hash function is picked for each message.

Before we consider which transmissions require authentication, it is important to realize
that any man-in-the-middle attack that results in differences between Bob's and Alice's
strings of key material can be detected by an equivalence check following error correction.
Alice and Bob perform the equivalence check by applying the same authentication hash
function to each of their strings and comparing the result.  If the results match, they
conclude that the strings are identical with a high probability.  This check is discussed
in more detail in the next section.  For the purposes of discussing authentication and
sifting, it is sufficient that we require Alice and Bob to perform the equivalence check
as a part of the protocol.

The transmissions on the quantum channel do not require authentication, since a
man-in-the-middle attack by Eve on the quantum channel will become evident
when the error correction process reveals that there is no correlation between
Alice's and Bob's sifted strings. (Note that the strings would also fail the equivalence
check.)

It is not strictly necessary for Alice and Bob to authenticate the classical messages they
exchange during sifting, since the equivalence check will eventually detect a
man-in-the-middle attack on the sifting protocol.  Nevertheless it {\it is\/}
advantageous for Alice and Bob to authenticate these messages.  Note that Eve
can compromise the secrecy of the final key by a man-in-the-middle attack only if she
makes the attack both on the quantum transmission and on the classical messages used for
sifting.  An attack on the classical messages alone results only in
a ``denial-of-service" attack, that is, an attack that makes it difficult or impossible
for Alice and Bob to complete the
protocol successfully, but that does not compromise the secrecy of the
key material.  Authentication of sifting renders these
attacks ineffective early in the protocol and thus localizes the attack
to the quantum transmission and sifting phases of the protocol.  This is
useful for Alice and Bob to know in formulating a response to the
attack. Authentication guarantees Alice and Bob that they are
working with the same subset of the pulses sent by Alice and that any
remaining errors are due to physical imperfections of the equipment or
attempts by Eve to measure, and therefore disturb, the pulses sent by Alice.
Any further interference by Eve is restricted to
denial-of-service attacks during error correction.

The authentication of the classical discussion results in a cost to the
overall rate of quantum key generation, since some of the secret bits
produced by previous iterations of the protocol must be sacrificed to
generate an authentication tag that Alice or Bob can validate but that
Eve cannot forge.  Wegman and Carter \cite{wc} show that
the size of the secret index required to select a hashing function is

\be
\label{E:hashkeysize}
w\left( g,c\right) = 4\left( g+\log_2 \log_2 c\right) \log_2 c
\ee

where $c$ is the length in bits of the message to be authenticated and $g$ is
the length in bits of the authentication tag.  The full and complete expression for the
quantity that we denote by $w$ and refer to as
the Wegman-Carter function, which is of crucial importance
in practical quantum cryptography, does not appear to have been properly analyzed
previously in the context of QC (nor apparently even {\it named}
by any authors). Surprisingly,
the closed-form function, as
such, doesn't appear as a numbered equation in \cite{wc}. In fact, it must be obtained
instead by combining quantities that appear in lines 3 and 17 in the first paragraph of
section 3 in \cite{wc}.  See \cite{gh} for a complete
discussion of the cost of authentication and its effect on the secrecy capacity
of the quantum key distribution system.

\subsection{Error Correction Phase}

At this point Bob and Alice move on to the error correction phase.  We will
estimate the
authentication, communication, and computational costs for a modified version
of the
error correction
protocol described by Bennett {\it et. al.\/}, \cite{bbbss}.  More
efficient techniques have been developed, for example the
``Shell" and ``Cascade" protocols described in \cite{brassardsalvail},
but the method described here is more suitable for
our purposes
since it is simpler to analyze.

At the beginning of the error correction phase, Alice and Bob each have a
string of $n$
bits.  The strings are expected to be nearly identical, but they will also
contain errors
for which Alice and Bob disagree on the value of the bit.  It is the goal of
error
correction to identify and remove all of these errors, so that Alice and Bob
can
proceed with a high degree of certainty that the strings are identical.
Error correction consists of three steps.  The first
step is the error detection and correction step, which eliminates all or
almost all
of the errors.  The validation step which follows eliminates any residual
errors and
iteratively tests randomly chosen subsets of the string to generate a high
degree of
confidence that the strings are identical.  The final step is authentication,
which
protects against a man-in-the-middle attack by Eve during the error correction
process.

At the beginning of the error detection and correction step, Alice and Bob
each
shuffle the bits in their string using a random shuffle upon which they have
previously
agreed.  The purpose of this shuffle is to separate bursts of errors so that
the
errors in the shuffled string are uniformly distributed.  Alice and Bob may
use
the same shuffle each time they process a new string of sifted bits, and
security is not compromised if Eve has complete prior knowledge of the
shuffling
algorithm, even including any random numbers used as parameters.

The error detection and correction step is an iterative process.
Alice and Bob begin each iteration $i$ by breaking their strings
into shorter blocks.  The block length is chosen so that the
expected number of errors in each block is given by a parameter
$\varrho$.  This is achieved by breaking the string into
$J^{\left( i\right)}$ blocks for the $i^{\rm th}$ iteration, where

\be
J^{\left( i\right)} = {\Bigg\lceil}{e_T^{\left( i-1\right)}\over\varrho}{\Bigg\rceil}~,
\ee

and $e_T^{\left( i\right)}$ is the expected number of errors
remaining after the $i^{\rm th}$ iteration or at the beginning of
the $i+1^{\rm st}$ iteration. In principle the parameter $\varrho
$ could change from iteration to iteration.  We assume that it is
a constant to simplify the analysis.  Alice and Bob compute the
parity of each of the blocks and exchange their results. Blocks
for which the parities do not match necessarily contain at least
one error. For each of the blocks in which Alice and Bob have
detected an error, they isolate the erroneous bit by a bisective
search, which proceeds as follows. Alice and Bob bisect one of the
blocks containing an error, that is, they divide it as evenly as
possible into 2 smaller blocks.  Alice and Bob each pick one of
the smaller blocks for the next parity check.  For definiteness,
assume they pick the block that lies closer to the beginning of
the shuffled string, which we will call the ``lower" block. The
other block is then the ``upper" block.  Alice and Bob then
compute the parity of the lower block and compare their results.
If the parities do not match, the error is in the lower block. If
they do match, the error is in the upper block. Alice and Bob then
bisect the block that contains the error and proceed recursively
until they find an erroneous bit. Bob then inverts that bit in his
string, and thus the error is removed.

We have described the bisective search as though the search were
completed for any block containing a detected error before beginning the
bisection on
the next block.  In fact, it is more efficient from a communications
standpoint to apply
each bisection to all the blocks with detected errors at the same time,
exchange parities
for all of the sub-blocks, and then to proceed recursively to the next
bisection.  This
results in fewer, but larger, packets of data for each exchange between Bob
and Alice,
thus reducing the overall frame overhead.

When the bisective search is completed for all blocks in which an error is
detected,
a new blocksize is computed based on the expected number of errors remaining,
the string
is broken up into a new set of larger blocks, parity checks are compared for
the blocks,
and bisective searches are made in those blocks containing detected errors.
This process is repeated until there would be only one or two blocks in the
string
for the next iteration, that is, until

\be
J^{\left( N_1+1\right)} \leq 2~,
\ee

where $N_1$ is the number of iterations in the error correction and detection
step.

In this version of the algorithm, the bits are not shuffled between successive
iterations of the error detection and correction step.  There is some value
in performing the shuffle, since it separates pairs of errors that may survive
previous iterations, thus making it more likely that they are found
before the validation step.  However, this step is not essential and has
not been included in the estimate of computational loads to follow.

The second step in the error correction phase, validation,
is also iterative.  During each iteration, Alice and Bob
select the same random subset of their blocks.  They compute the parities and
exchange
them.  If the parities do not match, Alice and Bob execute a bisective search
to
find and eliminate the error.  Iterations continue until $N_2$ consecutive
matching parities are found.  At this point, Alice and Bob conclude that their
strings
are error free.

The last step in the error correction phase is authentication.  Up until now,
Alice and Bob
have made no attempt to authenticate their exchange of parity information on
the classical
channel.  Eve could mount a man-in-the-middle attack during the error
correction phase
that would fool Alice and Bob into correcting the wrong set of bits.  This
would not give
Eve any additional information about the secret string, but it could result in
Alice and
Bob believing that their strings are identical when in fact they are not.
Even if one bit
is different, the privacy amplification phase will produce strings that are
completely
uncorrelated, and Alice and Bob will still believe that their strings are
identical.
The solution to this problem is for Alice and Bob to verify that their strings
are the same
at the end of the error correction phase.  This effectively authenticates
their prior
communications, since any successful attempt by Eve to steer the error
correction
process will be immediately apparent.

This approach presupposes that Alice and Bob can verify that their strings are
the same
without leaking too much additional information to Eve.  This can be
accomplished if
Alice and Bob apply the same hash function to their strings and compare the
resulting
tag.  This does not provide an absolute guarantee that the strings are the
same, but
if the hash function is chosen as described in \cite{wc}, the probability that two
different strings will yield the same tag is

\be
{\cal P}\left({\rm same~tag,~two~strings}\right) = 2^{-g_{EC}}~,
\ee

\noindent
where $g_{EC}$ is the length of the tag.  This gives a high degree of
confidence that
the strings are identical even for relatively short ($g_{EC}\sim 30$) tags.
The price Alice and Bob have to pay for this
is that they must use a portion of the secret bits obtained from previous
iterations
of the protocol to select the hash function, indicate whether the keys match,
and
authenticate their transmissions.

\subsection{Privacy Amplification Phase}

The general scheme of privacy amplification is described in \cite{bbcm} and
\cite{cw} .
The hash functions map a sifted, error corrected string of length $n$ to a
string of length $L_{pa}$,
where

\be
\label{E:palength}
L_{pa} \equiv n-e_T^{(0)}-q-t-\nu-g_{pa}~,
\ee

where $n$ is the length of the sifted string, $e_T^{(0)}$ is the number of errors removed
during error correction, $q$ is the additional information leaked during error
correction, $t$ is the amount of information Eve obtains by attacks on single photon
pulses, and $\nu$ is the amount of information she obtains by attacks on multi-photon
pulses.
The resulting string is thus shorter than the sifted string by the number of bits
that Eve
may have obtained by listening to the classical discussion, plus an additional
security parameter $g_{pa}$.
As shown in \cite{bbcm}, this parameter determines an upper bound on the
expected amount of information, $I$, that Eve can retain
following privacy amplification:

\be
\label{E:infopa}
\langle I({\tilde K}, X) \rangle \leq {{2^{-g_{pa}}}\over{\ln 2}}~,
\ee

where ${\tilde K}$ is the key after privacy amplification and $X$ is the information
Eve has obtained from all phases of the protocol.  The expectation value is
over the set of functions from which Alice and Bob choose their hash function.
See \cite{lutkenhaus-practical,ght} for a discussion of secrecy bounds
that are not
conditioned on an average over the hash functions.

Hash functions appropriate for privacy amplification are described by Carter
and Wegman
\cite{cw}.  The class
of hash functions used for authentication and equivalence checking is not
practical for privacy amplification due to
the much larger size of the output string.  The authentication hash functions
are designed to produce output strings that are no more than half as long as the input
string.  Since we wish to retain as much information as possible, it is clearly
advantageous to use hash functions that can produce an output string that
is nearly as long as the input
string.  Furthermore, recall that the length of the
index for choosing
an authentication hash function is \cite{wc, gh}
\be
w\left( g,c\right) = 4\left( g+\log_2 \log_2 c\right) \log_2 c
\ee
where $c$ and $g$ are the lengths of the input and output strings,
respectively.  For purposes
of authentication and error correction, an output string of length $g \leq 50$
is adequate,
and the length of the index is relatively short even for long input strings
due to the
logarithmic factors.  In privacy amplification, where the output
string is
nearly as long as the input string, this index is roughly 4
times as
long as the string to be hashed.  In contrast, the
hash functions suitable for privacy
amplification
are described by
two parameters, each as long as the input string, so that the total size of
the index is only twice as long as the input string.
The Carter-Wegman functions described
in \cite{cw} are therefore a much better choice for privacy amplification since they are
capable of producing keys nearly as long as the input and since they require
shorter indices for their definition than do the Wegman-Carter functions
given the large size of the output strings.

The error correction phase guarantees that the strings Alice and
Bob have obtained are identical with high probability.  Bob and
Alice implement privacy amplification by agreeing on an index and
applying the hash functions separately to their strings. The
resulting strings are identical and secret in the sense of privacy
amplification ({\it cf} eq.(\ref{E:infopa})). Note that the
sifting protocol itself supplies random strings of sufficient
length to define the required hash index.  Bob's choice of basis
for the $2n$ pulses he receives is one such source. Another
alternative is to compute the parities of the indices Bob sends to
Alice by which he identifies which pulses were detected by his
equipment.

\subsection{Computational Load}

The total computational load implied by the protocols described here is
analyzed in detail in \cite{gh}.  The result of the analysis is an
approximate upper bound on the number of instructions per block of key material:

\newpage

\bea
\label{E:compload}
{\cal L}_B &\leq&
   {\cal L}_B^{\left( 0 \right)} \nonumber\\
  &&+ \left( 50 + {220 \over g_{auth}} \right) n \left(1+ \log_2 m \right) \nonumber\\
  &&+ {\Bigg [} 200 + 25 N_1 +
            12.5 \left( 1 - e^{-2\varrho} \right) N_1 +
            25 \varrho +
            37.5 \left( N_2^{\left( n \right)} + N_2^{\left( f \right)} \right)\nonumber\\
     &&\qquad+{43 \over w} + {220 \over g_{auth}} + {110 \over g_{EC}} {\Bigg ]} n
\nonumber\\
  &&+ {46 \over {w^2}} n^2 ~.
\eea

In this expression, ${\cal L}_B^{\left( 0 \right)}$ is the ``non-iterative" portion of the
load, representing code that executes once for each block of data without iterating
bit-by-bit through the string of key material.  $m$ is the block size in bits of the raw
key
material sent by Alice to Bob over the quantum channel.  $n$ is the block size of the
sifted key material.  $g_{auth}$ and $g_{EC}$ are the security parameters for
authentication
and error correction, respectively.  $\varrho$ is the parameter
that determines the blocksizes used in successive iterations of the first step of
error correction.  $N_1$ is
the number of iterations required in the first step of error correction.  The term
$N_2^{\left( n \right)} + N_2^{\left( f \right)}$ is the total number of iterations
in the second step of error correction.  Expressions for these iteration counts in
terms of more fundamental parameters are found in \cite{gh}.  Finally, $w$ is the wordsize
in bits of the processing element that performs elementary integer arithmetic.

It is instructive to evaluate this expression for a practical example.
Assume a substantial non-iterative contribution:

\be
{\cal L}_B^{\left( 0 \right)} = 10^6 {\rm ~operations~per~block}~,
\ee

and take the wordsize of the processor to be 64 bits.  The other parameters are
chosen to have reasonable operational values and to give a reasonable
value for the computation rate as computed below
($m=2\times 10^8$ bits, $n=2\times 10^5$ bits,
$e_T^{\left( 0\right)} = 2 \times 10^3$ bits, $\varrho = 0.5$,
$g_{EC} = g_{auth} = N_2 = 30$).
The resulting estimate of the load is 1.1 billion operations per block.   The quadratic
term contributes 450 million operations to the total.  Of the other terms, the dominant
contributions are the term in in $N_2^{\left( n \right)} + N_2^{\left( f \right)}$,
which is due to parity checks and random
block extractions during the validation step of error correction, and the term in
$\left( 1+\log_2 m\right)$, which is due to sifting.
Note that the non-iterative overhead load is negligible in comparison with the other
contributions.  This indicates that a substantial amount of ``bookkeeping" code can
be included along with the core software that is essential to arriving at the final
secret key without significantly affecting the processing requirements.  One of the uses
of eq.(\ref{E:compload}) is to establish a load budget for such code
during software design and implementation
to ensure that the bulk of the
processing resources are available for the core software functions.

The computation rate ${\cal R}_B^{comp}$ required to support key distribution
is found by dividing the load per block by
the time required to transmit one block over the quantum channel:

\be
{\cal R}_B^{comp} = {{\cal L}_B\over m\tau}~,
\ee

where $m$ is the raw block size, and $\tau$ is the bit cell period for sending each bit.
Note that the processing load is quadratic in the sifted block size, so that the
computation rate increases roughly linearly with the block size.  This means that
the computations required for the full protocol imply upper bounds on the sifted block
size $n$ and, by extension, the raw block size $m$
in order to keep the computational load within
the capabilities of the equipment.

In order to obtain a quantitative result for the computational rate, we must choose a
value
for $\tau$, which is the inverse of the pulse repetition frequency of the photon
source.  Experimental demonstrations of quantum key distribution have
used
attenuated lasers with pulse rates up to 1 MHz \cite{demorefs1}, and photon transmission
and detection has been demonstrated at 400 MHz \cite{demorefs6}.
As discussed in \cite{gh}, the
limiting physical
factors include the switching speed of the optical equipment and the time required
for the single photon detectors to reset for the next pulse.  A rate of 10 GHz is
feasible with high quality optoelectronics, but current photon detectors are much too
slow to support this.  It is possible that new technologies \cite{sobopapers} will enable
single photon detectors to support these rates.  We will accordingly set
$\tau = 10^{-10}$ sec.  The computation rate for our example is then

\be
\label{56billion}
{\cal R}_B^{comp} = 56 {\rm ~billion~operations/sec}~.
\ee

This is rather high for a single general purpose
processor, but should be achievable in a parallel architecture in which each block
of the input data is allocated to a single processor as it becomes available.
Recall also that general purpose computers are far from optimal for this type of operation.
Most of the processing steps involving the packing and unpacking of the bits would not be
necessary in a special purpose device, and many of the other processing steps,
notably block parity
computations and random selection of substrings, could be accomplished much more
efficiently using special purpose hardware.  In any case, it should be clear that
computational power is not a limitation on the practicality of quantum cryptography
even at very high rates of operation.

\section{Effective Secrecy Capacity and Rate of Key Generation}

In this section we apply the analysis key generation rate analysis
of \cite{gh} to the consideration of
system performance characteristics in representative
practical implementations
of quantum cryptography for which optical fiber is used as the quantum channel.
In particular we wish
to compare the use of perfect single photon sources with the use of weak coherent sources,
and we consider this in distinct
scenarios in which the eavesdropping attacks are
constrained to lesser and greater degrees.

The motivation for this discussion is the common misconception that pure single photon
sources inherently provide a higher degree of secrecy than do sources with some admixture
of multi-photon pulses.  This misconception arises from the following argument.  If Eve
measures single photon states, she necessarily introduces a disturbance that can be
detected by Alice and Bob. The secrecy of the key is thus protected by the laws of
quantum mechanics.  However, if some of the pulses encoding the key contain multiple
photons, Eve can measure a fraction of the photons in any given multi-photon pulse without
disturbing the rest, thus obtaining information without being detected.  This appears to
circumvent the guarantee of secrecy for single photon sources.  The need to
carry out privacy amplification, however, applies whether the protocol is
implemented with a pure single photon source {\it or\/}
with a source that produces
both single and multiple photon pulses. Even if there are no multi-photon
pulses {\it at all} amongst
the signals sent from Alice to Bob, the fact that the physical hardware generates
errors, combined with the fact that Eve may be present,
results in the need for privacy amplification. {\it Precisely the same
degree of secrecy} is realized whether
the protocol is implemented with pulsed lasers
generating weak coherent pulses or with single photon sources.  The chief difference is
the rate of generation of key material.  In principle the rates
achievable with single photon sources should be higher, all other things being equal,
since Eve cannot gain information from attacks on
multi-photon pulses in this case.  In reality, none of the single photon sources
now in existence
can produce pulses at a rate comparable with currently available pulsed lasers.
In this section we show that the key generation rates are much higher for weak
coherent sources ({\it i.e.\/} for sources producing some fraction of multi-photon
pulses) than for single photon sources for implementations that are likely to be
feasible in the near future.

We first summarize the results of \cite{gh} for key generation
rates achieved with weak coherent sources.  We next present
results suitably modified for ideal single photon sources.
Following \cite{gh}, we consider only individual attacks, that is,
attacks in which Eve attempts to obtain information by making
measurements of individual photons.  The case of a more general
quantum attack, in which Eve may entangle probes with arbitrarily
chosen groups of photons is a subject for a subsequent analysis.

\subsection{Definitions}

We define the secrecy capacity ${\cal S}$ as
the ratio of the length of the final key to the length of the original string of pulses
sent from Alice to Bob over the quantum channel:
\be
{\cal S} = {L \over m}~.
\ee
This quantity is useful for two reasons.  First, it can be used in proving the
secrecy of specific practical quantum cryptographic protocols by establishing that
the inequality
\be
{\cal S} > 0
\ee
holds for the protocol.  Second, it can be used to establish the rate of generation
of key material according to
\be
{\cal R} = {{\cal S} \over\tau}~,
\ee
where $\tau$ is the pulse period of the initial sequence of photon transmissions. We refer
to ${\cal R}$ as the effective secrecy rate.

The length of the final key is given by
\be
L = n - \left( e_T + q + t + \nu \right) - \left( a + g_{pa} \right)~,
\ee
which is the same as the length of the key after privacy amplification ({\it cf}
eq.(\ref{E:palength}))
except that an additional amount $a$ has
been subtracted to account for the secret bits required for authentication
during the next iteration of the protocol. Recall that $n$ is the length of the sifted
string, $e_T \equiv e_T^{(0)}$ is the number of errors removed
during error correction, $q$ is the additional information leaked during error
correction, $t$ is the amount of information Eve obtains by attacks on single photon
pulses, and $\nu$ is the amount of information she obtains by attacks on multi-photon
pulses.
See \cite{gh, ghprla} for a detailed
quantitative treatment of the terms appearing in this result.  We apply this
analysis to a number of scenarios involving key distribution system operating
over a fiber optic quantum channel in section \ref{S:scenario}.

\subsection{Effective Secrecy Rates}
\label{S:scenario}

There is extensive research activity \cite{sps_exp1,sps_exp2,sps_exp3,sps_exp4,sps_exp5}
currently taking place devoted to the development of sources of single photons to serve
as quantum bits in various applications, including in quantum cryptography. At the present
time, though, there are no robust, {\it perfect} sources of single
photons available that can be used for practical quantum cryptography purposes.
At the same time a number of groups have carried out
demonstrations of quantum key distribution in which filtered, pulsed lasers are used
as the source of the quantum bits. The use of such weak coherent sources results in
the production of both single-photon pulses and multiple-photon pulses
in the transmission stream from Alice to Bob. The presence of multi-photon
pulses allows Eve
to execute a set of attacks that may (depending on precisely how Bob monitors
his photon detector) require a significant amount of
privacy amplification compression in order to assure secrecy, resulting in a considerably
reduced throughput rate. However, under the conditions described in
Section \ref{S:ss} above, privacy amplification
is also required in any practical implementation even if we utilize a perfect single
photon source. Thus with the use of either type of source we see that the final,
shared key is characterized by unconditional secrecy in the sense of privacy
amplification ({\it cf} eq.(\ref{uspa})). Then there is no secrecy advantage
that inures to the use of a single photon source, and the only figure-of-merit that
distinguishes between the use of a single photon source (SPS) and a weak coherent
source (WCS) is the effective secrecy {\it rate}.

To analyze and compare the effective secrecy rates due to single photon sources
and weak coherent sources we must employ the
appropriate expression for the number of sifted bits shared between Alice and Bob
in each case. This is given for a WCS by \cite{gh}
\bea
\label{nwcs}
n_{WCS} &=& {m\over 2}{\Bigg [}\left(1-r_d\right)\psi_{\ge 1}\left(\eta\mu\alpha\right)
+r_d{\Bigg ]}
\nonumber\\
&\simeq&{m\over 2}{\Bigg [}\psi_{\ge 1}\left(\eta\mu\alpha\right)+r_d{\Bigg ]}~,
\label{aaa}
\eea
where $\mu$ is the average number of photons per pulse, $\eta$ is the
efficiency of Bob's detector, $\alpha$ is the transmission
probability in the quantum channel, and $r_d$ is the probability of obtaining a
dark count in Bob's detector during a single pulse period.
$\psi_{\ge 1}\left(X\right)$ is
the probability of encountering $1$ or more photons in a pulse selected at random from
a stream of Poisson pulses having a mean of X photons per pulse:
\be
\label{E:psi}
\psi_{\ge 1}\left(X\right)=\sum_{l=1}^\infty e^{-X}{X^l\over l!},
\ee
and we have assumed that $r_d<<1$ in the second line in eq.(\ref{nwcs}).
The corresponding expression that arises in the case
of a {\it perfect} source of single photon quantum bits is \cite{ghprlb}
\bea
n_{SPS} &=& {m\over 2}{\Bigg [}\left(1-r_d\right)\eta\alpha+r_d{\Bigg ]}
\nonumber\\
&\simeq&  {m\over 2}\left(\eta\alpha+r_d\right)~.
\label{111}
\eea
As expected, we see that $n_{SPS}$ is independent of the quantity $\mu$. The
corresponding expressions for the numbers of errors in the sifted strings generated
by the two types of sources are
\be
e_{T,WCS} \simeq
{m\over 2}{\Bigg [}\psi_{\ge 1}\left(\eta\mu\alpha\right) r_c
+{r_d \over 2}{\Bigg ]}~,
\ee

($r_c$ is the intrinsic channel error fraction) and

\be
e_{T,SPS} \simeq
{m\over 2}\left(\eta\alpha r_c+{r_d \over 2}\right)~.
\ee

We recall that $q$ and $t$ are, respectively, the information leaked during error
correction and an upper bound for the amount of information Eve can obtain by direct
measurement on single photon pulses. Upon introducing the quantities $Q$ and $T$ through
\be
q\equiv Qe_T~,
\ee
and
\be
t\equiv Te_T~,
\ee
the expressions for the effective secrecy capacities for the two types of
sources can be shown to be given by \cite{gh,ghprla,ghprlb}

\bea
\label{162}
{\cal S}_{WCS}&\equiv&{n - e_T - q - t - \nu -g_{pa}-a\over m}{\Big\vert}_{WCS}
\nonumber\\
&=&{n-fe_T-\nu-g_{pa}-a\over m}{\Big\vert}_{WCS}
\nonumber\\
&=&{1\over 2}{\Bigg [}\psi_{\ge 1}\cdot\left(1-f_{WCS}r_c\right)+\left(1-{f_{WCS}\over 2}
\right)r_d-\tilde\nu{\Bigg ]}-{g_{pa}+a\over m}~,
\eea

(here $\psi_{\ge 1}\equiv\psi_{\ge 1}\left(\eta\mu\alpha\right)$ and
$\tilde\nu\equiv 2\nu /m$ with the rescaled quantity $\tilde\nu$ independent of $m$)
and

\bea
\label{xxx}
{\cal S}_{SPS}&\equiv&{n-e_T-q-t-g_{pa}-a\over m}{\Big\vert}_{SPS}
\nonumber\\
&=&{n-fe_T-g_{pa}-a\over m}{\Big\vert}_{SPS}
\nonumber\\
&=&{1\over 2}{\Bigg [}\eta\alpha\cdot\left(1-f_{SPS}r_c\right)+\left(1-{f_{SPS}
\over 2}\right)r_d{\Bigg ]}-{g_{pa}+a\over m}~,
\eea

where we have introduced
\be
f_{WCS}\equiv 1+Q_{WCS}+T_{WCS}
\ee
and
\be
f_{SPS}\equiv 1+Q_{SPS}+T_{SPS}~,
\ee
with $Q_{SPS}$ calculated solely as a function of $n_{SPS}$ and
$e_{T,SPS}$, and with $Q_{WCS}$ calculated solely as a function of
$n_{WCS}$ and $e_{T,WCS}$, respectively (and likewise for the
calculation of $T_{WCS}$ and $T_{SPS}$). Detailed expressions for
$Q$ and $T$ may be found in \cite{gh,ghprla,ghprlb}.

We now proceed to compare the use of a perfect single photon source with the use of
a weak coherent source. We also wish to consider two levels of technology
available to the enemy, one in which the enemy can surreptitiously effectively
eliminate the intrinsic attenuation along the quantum channel, which is usually
to her advantage, and another in which
this is not possible for her to do. In the former case we may imagine two ways
in which, at least in principle, it could be possible for this to occur. In one situation,
we may imagine that the
enemy somehow has
the capability to remove the installed fiber optic cable and replace it with
a new cable that is effectively lossless, all without being detected ``in the act." Of
course, it is clear that this is an extremely unlikely situation, due both to the
near-impossibility of undetectably
removing and replacing a long length of cable, and, even
moreso, the implausibility of being
able to produce lossless cable at all. We will underscore this implausibility by
referring to such a hypothetical lossless
fiber optic cable as ``magic" cable. The other situation in which the enemy may effectively
eliminate the attenuation along the quantum channel arises if she has both an
accomplice near Bob and access to prior shared entanglement. In that case
Eve and the accomplice located near Bob prepare pairs
of entangled photons in advance. Eve then entangles one of these pairs
with a photon emitted by Alice. Her accomplice can then make measurements on the entangled
state, gaining information about the photons at Eve's location without losing photons
to the attenuation in the channel.

For the calculation of ${\cal S}_{WCS}$ and ${\cal R}_{WCS}$ we will need
to use the correct
expression for $\nu$, the function that specifies the amount of privacy amplification
compression that is associated to the presence of multi-photon pulses in the transmission
from Alice to Bob. The appropriate form for $\nu$ is determined by whether or not we
presume that it is possible for Eve to effectively eliminate the line attenuation along
the quantum channel, as described above. As discussed in detail in \cite{gh,ghprla}, the
correct expressions for $\nu$ that will ensure unconditional secrecy (in the sense of
privacy amplification) are given follows: If we presume that Eve {\it can} somehow
effectively eliminate the line attenuation along the quantum channel we have

\be
\label{region1}
\nu^{max}={m\over 2}\llb \psi_{\ge 2}\left(\mu\right)
-\left(1-y\right)^{-1}\cdot\;{\Bigg \{}e^{-y\mu}-e^{-\mu}
{\Bigg [}1+\mu\left(1-y\right){\Bigg ]}{\Bigg \}}\rrb~,
\ee

with the parameter $y$ given by $y=\eta$, subject to the constraint
$y~>~1-{1\over\sqrt 2}~~{\Big (}i.e.,~y~\gsim~0.293{\Big )}$, which is
automatically satisfied in
the examples we consider. If we presume that Eve {\it cannot}
effectively eliminate the line attenuation along the quantum channel we have
\be
\label{region2}
\nu^{max}={m\over 2}{\Bigg [}\psi_2\left(\mu\right)y+1
-e^{-\mu}{\Bigg (}\sqrt 2\sinh{\mu\over\sqrt 2}+2\cosh{\mu\over\sqrt 2}-1{\Bigg )}
{\Bigg ]}~,
\ee
with the parameter $y$ now given by $y=\eta\alpha$, subject to the constraint
$y~<~1-{1\over\sqrt[3] 2}~~{\Big (}i.e.,~y~
\lsim~0.206{\Big )}$, which is also automatically satisfied in the examples we consider.

In the scenarios that follow we consider
fiber-optic cable implementations of quantum key distribution
making use of either good quality single-mode,
polarization-preserving fiber characterized by an intrinsic attenuation characteristic
of $A_1=0.3~{\rm dB}$ per kilometer, or of high quality fiber characterized by
$A_2=0.2~{\rm dB}$ per kilometer. We take the photon detector device efficiency to be
$\eta=50\%$, and we assume that appropriate splicing and insertion of suitable
dispersion-compensating fiber segments, as discussed in \cite{gh}, has been carried out
so as to mitigate the dispersion losses described and
analyzed there. To account for the
associated splicing loss and other insertion losses we assume that the quantum
channel is characterized by a total bulk loss of $\kappa=-5~{\rm dB}$, in addition to the
losses associated with the attenuation per unit length. Thus the transmission
probability for the quantum channel is given by
\bea
\label{205}
\alpha&=&\alpha\left(L_{fiber},A,\kappa\right)
\nonumber\\
&=&10^{-{AL_{fiber}+\kappa\over 10}},
\eea
where $L_{fiber}$ is the length of the fiber cable connecting Alice and Bob.

For all of our numerical examples we have taken a value for the Shannon
deficit parameter $x$ of $x=1.16$ ({\it cf} \cite{gh} for a discussion of $x$),
which means that we are assuming that an efficient method of error
correction has been employed that approaches the Shannon limit to within 16\%, and
we use a raw bit processing block size of $m=200$ Megabits. In addition, we have
also set all of the continuous authentication security parameter values, $g_i$,
as well as the privacy amplification security parameter
$g_{pa}$, equal to 30, and we have employed a value of $\epsilon=10^{-9}$ for the
selectable infinitesimal quantity that determines the success likelihood for attacks
on single-photon pulses, (this applies for both
authentic single photon qubits generated by a single
photon source, and the single-photon
{\it part} of the transmission stream generated by a filtered, pulsed laser). We are also
assuming the use of a photon detector device that has a dark count rating of $10^{-6}$
counts per bit cell. In addition, we make the further assumption
that the intrinsic channel error rate due to pulse dispersion in
the optical fiber is no greater than 1\%.

It is important to point out a distinguishing feature in the computations of the
effective secrecy capacities for the
WCS and SPS cases. In calculating ${\cal S}_{WCS}$ we want to determine the value of
$\mu$ that produces the maximum throughput, whereas in the calculation
of ${\cal S}_{SPS}$
this issue doesn't arise since each pulse carries precisely one photon.
When we plot ${\cal R}_{WCS}$ as a function of position along the
fiber-optic quantum channel we utilize this optimized value
of the mean photon number per pulse, $\mu_{{\rm opt}}$,
and in particular we must do so for {\it each point
along the path}. This is because $\alpha\left(L_{fiber},A,\kappa
\right)$ is explicitly a function of $L_{fiber}$, the position of Bob on the quantum
channel ({\it cf} eq.(\ref{205})). As
the effective secrecy capacity ${\cal S}_{WCS}$ itself depends on $\alpha$, we must
determine
new values of $\mu_{{\rm opt}}$ for each value of $L_{fiber}$ by obtaining a new
solution of the optimization
equation $0=\partial_\mu {\cal S}_{WCS}\vert_{\mu=\mu_{{\rm opt}}}$.
In contrast, ${\cal R}_{SPS}$ and ${\cal S}_{SPS}$ have no $\mu$-dependence at all.

\subsubsection{Scenario One: Eavesdropper {\it {\bf cannot}} eliminate line attenuation}

\begin{figure}[htb]
\vbox{
\hfil
\scalebox{0.7}{\rotatebox{0}{\includegraphics{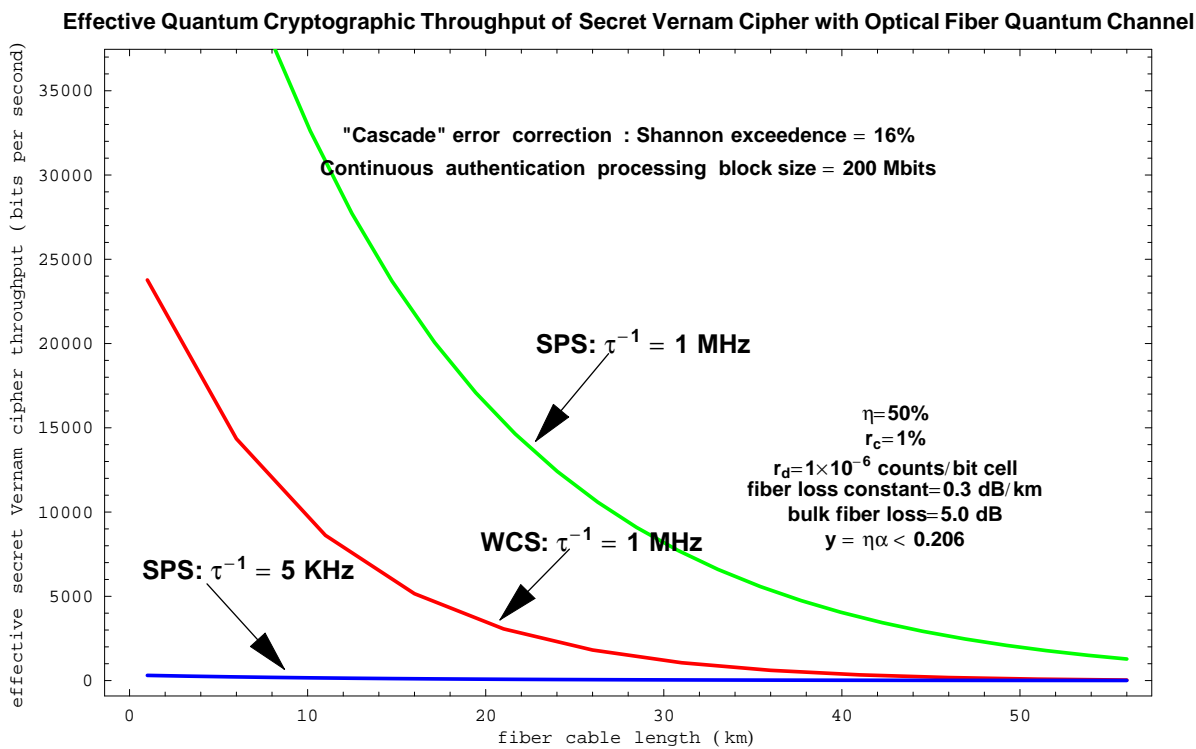}}}
\hfil
\hbox to 0.25in{\ }
}
\bigskip
\caption{%
Effective Rate Curves for Fiber-Optic Link {\it without} Surreptitious Cable Replacement
or use of Prior Shared Entanglement by Eve
}
\label{F:algofig2}
\end{figure}

In Figure \ref{F:algofig2} we plot three effective secrecy rate curves. In this example,
we {\it presume} that the enemy can neither (a) surreptitiously replace the installed
cable with a different cable
(in particular this means that the enemy is presumed to be unable to surreptitiously
replace the installed cable with an effectively lossless, or ``magic" cable),
nor (b) make use of prior shared entanglement as a
resource for cryptanalytic attacks. This means that the enemy cannot alter the
actual value of the line attenuation that exists along the cable. In turn, this
means that \cite{gh} we should set the privacy amplification parameter $y$ to
$y=\eta\alpha$ in the {\it multi-photon part} of the privacy amplification function to
be used in the calculation of ${\cal S}_{WCS}$ and ${\cal R}_{WCS}$. There is
no analogous issue in the case of the calculation of $S_{SPS}$ and ${\cal R}_{SPS}$,
since there is no multi-photon term in the privacy amplification function in the case
of a single photon source. Thus, for the computation
of ${\cal S}_{WCS}$ and ${\cal R}_{WCS}$
we make use of the form of $\nu^{max}$ given in eq.(\ref{region2}) above.

The upper
curve displays the effective secrecy rate that would arise with a perfect single photon
source operating at a pulse repetition frequency (PRF) of 1 MHz. The middle curve gives
the
effective secrecy rate corresponding to the use of a weak coherent source, also
operating at a pulse repetition frequency of 1 MHz. In order to represent a more realistic
situation based on the current state of the art in perfect single photon generation,
the lower curve shows the rate that
is obtained with the use of a perfect single photon source operating at a pulse repetition
frequency of 5 KHz. Assuming that we compare systems with equal pulse repetition
frequencies, there is a substantial gain realized with the use of a perfect single photon
source compared to the use of a weak coherent source. For instance, inspection of the
graph reveals that at a separation distance of $L_{fiber}=$
10 kilometers an effective secrecy rate of
about 9840 bits per second can be realized with the use of a weak coherent source with
a 1 MHz PRF. A
perfect single photon source operating at a PRF of 1 MHz achieves this same secrecy
rate at a distance of about 27.5 kilometers. We may also compare rates between the
two systems at a fixed separation distance between Alice and Bob. For example, at a
separation distance of 10 kilometers, in going from a WCS system to a SPS system
the rate increases from 9840 bits per second to
32900 bits per second, a gain of about 5.2 dB.

\begin{figure}[htb]
\vbox{
\hfil
\scalebox{0.7}{\rotatebox{0}{\includegraphics{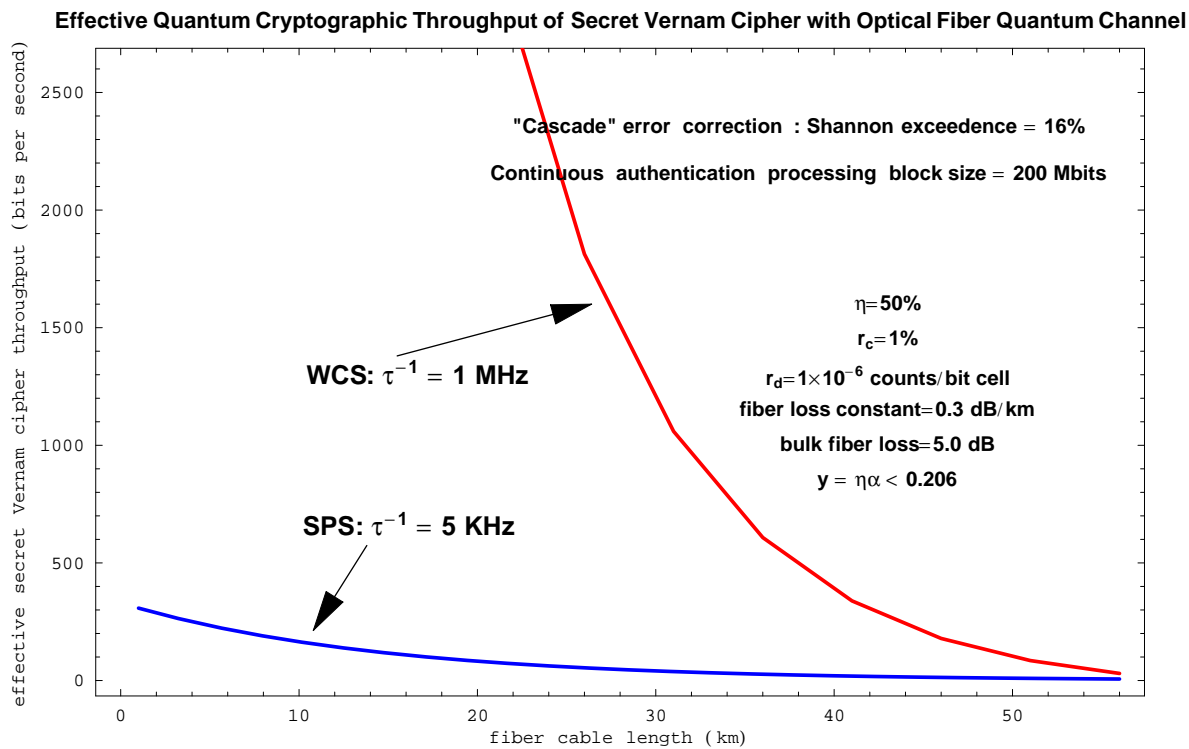}}}
\hfil
\hbox to 0.25in{\ }
}
\bigskip
\caption{%
Effective Rate Curves for Fiber-Optic Link {\it without} Surreptitious Cable Replacement
or use of Prior Shared Entanglement by Eve
}
\label{F:algofig3}
\end{figure}

Although on the scale of the graph in Figure \ref{F:algofig2} the curve for the SPS
system operating at a PRF of 5 KHz
appears to yield no throughput at all, inspection of Figure \ref{F:algofig3}
reveals that this is not the case. We note that, in spite of the {\it substantial}
reduction
in the amount of privacy amplification compression that is realized upon
going from a WCS system
to a SPS system ({\it i.e.}, replacing $\nu^{max}$ with 0),
there is no location, up to a separation distance of 56 kilometers, at
which the 1 MHz WCS curve and 5 KHz SPS curve cross each other. Therefore, if we
take a putative 5 KHz SPS system as representative of what might be achieved
in the near future, we see that it is nevertheless advantageous to employ a 1 MHz
WCS system compared to the 5 KHz SPS system.

\subsubsection{Scenario Two: Eavesdropper can eliminate line attenuation}

\begin{figure}[htb]
\vbox{
\hfil
\scalebox{0.7}{\rotatebox{0}{\includegraphics{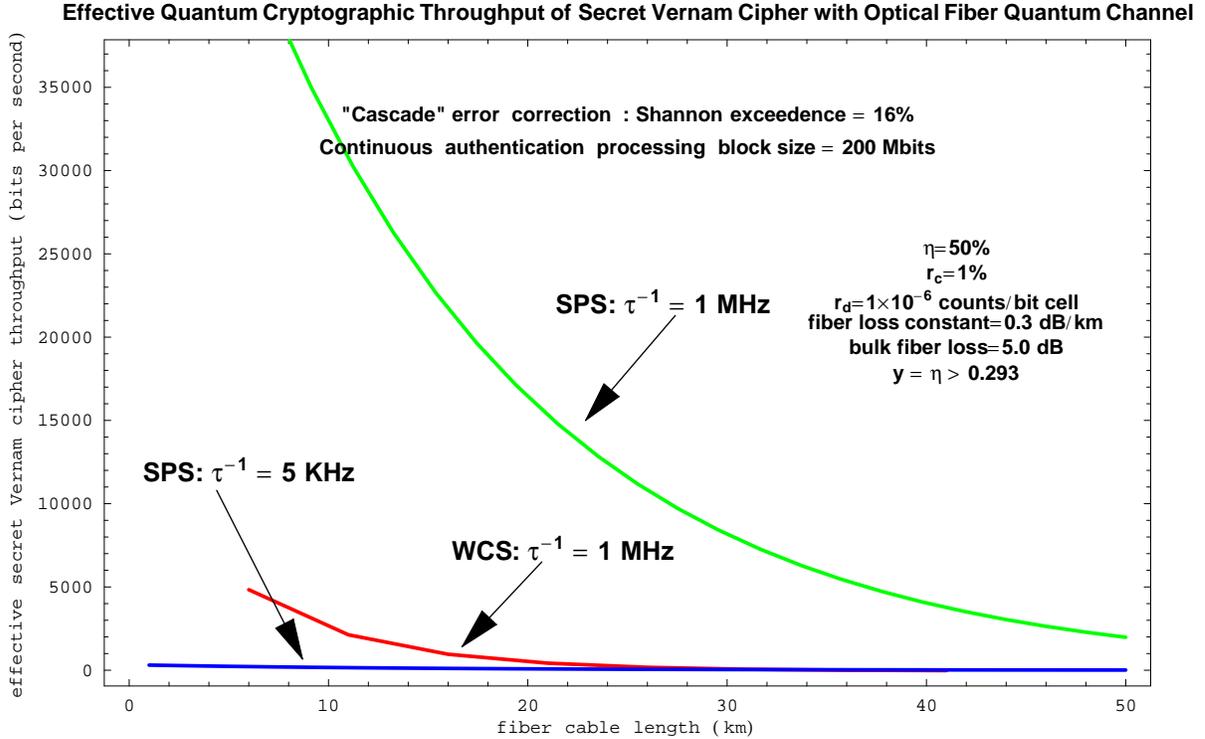}}}
\hfil
\hbox to 0.25in{\ }
}
\bigskip
\caption{%
Effective Rate Curves for Fiber-Optic Link {\it with} Surreptitious Cable Replacement
or use of Prior Shared Entanglement by Eve
}
\label{F:algofig5}
\end{figure}

In Figure \ref{F:algofig5} we consider a scenario in which the enemy {\it is}
somehow able to effectively
eliminate the attenuation along the quantum channel, either by surreptitiously
replacing the installed cable with a ``magic" cable that is lossless, or by employing
prior shared entanglement distributed between two operating locations adjacent to the
Alice and Bob locations (but somehow unobserved by Alice and Bob).

As in Figure \ref{F:algofig2}, the upper
curve in Figure \ref{F:algofig5} displays
the effective secrecy rate that would arise with a perfect single photon
source operating at a pulse repetition frequency of 1 MHz. The middle curve gives the
effective secrecy rate corresponding to the use of a weak coherent source, also
operating at a pulse repetition frequency of 1 MHz. The lower curve shows the rate that
would arise with the use of a perfect single photon source operating at a pulse
repetition
frequency of 5 KHz. Assuming as before
that we compare systems with equal pulse repetition
frequencies, we see that
there is a substantial gain realized with the use of a perfect single photon
source compared to the use of a weak coherent source. Inspection of the
graph reveals that at a separation distance of $L_{fiber}=$ 10 kilometers one obtains an
effective secrecy rate of
about 2130 bits per second with the use of a weak coherent source operating at a PRF
of 1 MHz. A
perfect single photon source also
operating at a PRF of 1 MHz achieves the same effective secrecy
rate at a distance of about 50 kilometers, a substantial increase.
We may also compare rates between the
two systems at a fixed separation distance between Alice and Bob. For example, at a
separation distance of 10 kilometers, upon going from a WCS system to a SPS system
the rate increases from 2130 bits per second to
30740 bits per second, a gain of about 11.6 dB.

\begin{figure}[htb]
\vbox{
\hfil
\scalebox{0.7}{\rotatebox{0}{\includegraphics{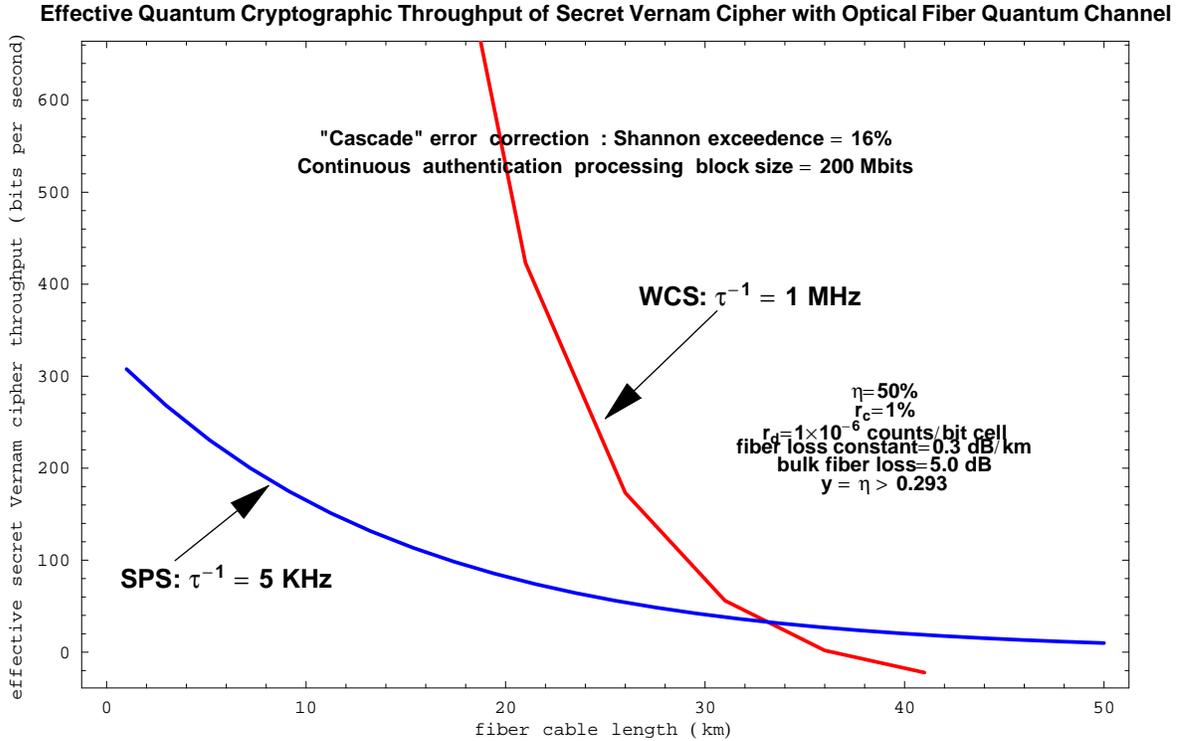}}}
\hfil
\hbox to 0.25in{\ }
}
\bigskip
\caption{%
Effective Rate Curves for Fiber-Optic Link {\it with} Surreptitious Cable Replacement
or use of Prior Shared Entanglement by Eve
}
\label{F:algofig6}
\end{figure}

As with the example given in Scenario One above, the
scale employed in Figure \ref{F:algofig5} makes it difficult to directly compare the
cases of a WCS source operating at a PRF of 1 MHz and a SPS source operating at
a PRF of 5 KHz.
Inspection of Figure \ref{F:algofig6}, however, reveals an interesting feature that
distinguishes Scenario Two from Scenario One. For
separation distances of less than 33 kilometers it is apparent that it is always
preferable
to employ a 1 MHz WCS source rather than a 5 KHz SPS source, as the throughput rate for
the former is always greater than that for the latter. However, the two throughput
curves cross each other at a separation distance of about 33 kilometers. Thus, for
distances greater than about 33 kilometers we find that one obtains a better secrecy
rate by using a SPS source instead of a WCS source, even though in this case the PRF of
the WCS source is 200 times larger than the PRF of the SPS source.

\subsubsection{Scenario Three: Comparison of Optical Fiber Quality}

\begin{figure}[htb]
\vbox{
\hfil
\scalebox{0.7}{\rotatebox{0}{\includegraphics{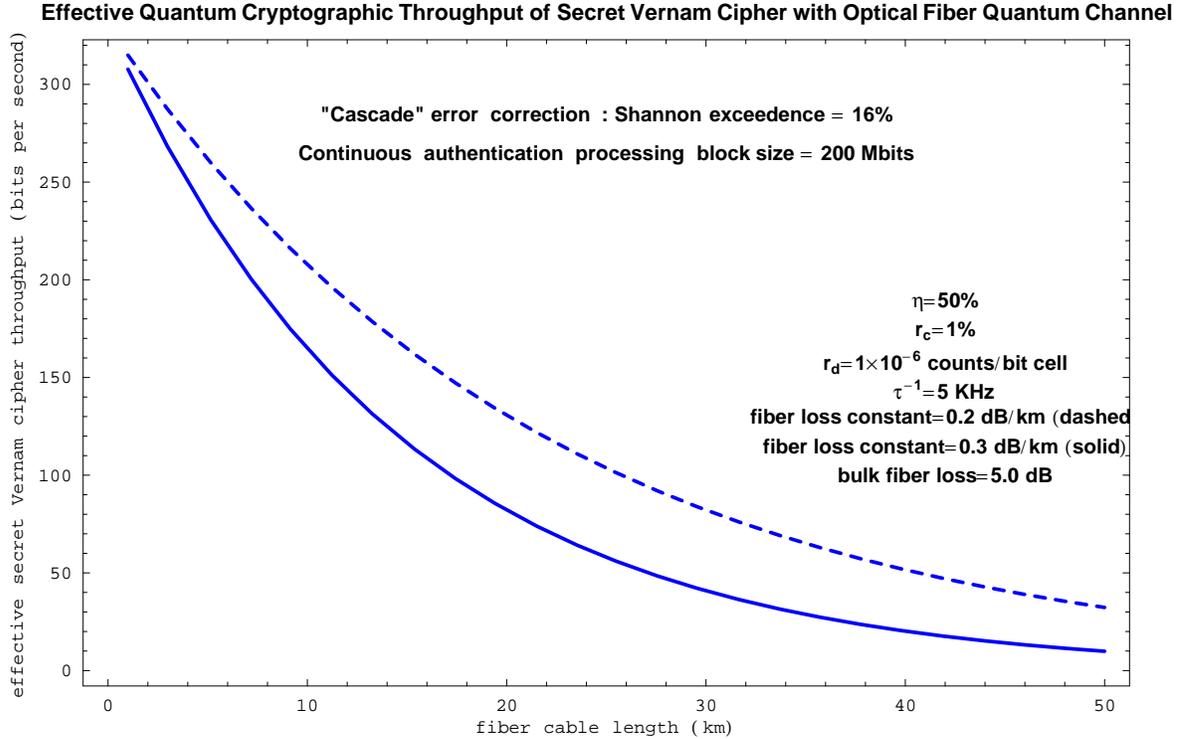}}}
\hfil
\hbox to 0.25in{\ }
}
\bigskip
\caption{%
Effective throughput Rate Comparison for Single Photon Source with Fiber Cables
of Two Different Qualities
}
\label{F:algofig4}
\end{figure}

In Figure \ref{F:algofig4} we illustrate the effect of improving the intrinsic
attentuation characteristic of the optical fiber, demonstrated here for the case
of a SPS source operating at a PRF of 5 KHz. The two curves in the graph are plotted
for two different fiber-optic
cables, with intrinsic attenuation values of 0.2 dB per kilometer
and 0.3 dB per kilometer, respectively, in the upper and lower curves.

Inspection of the graph reveals that, at a separation distance between
Alice and Bob of $L_{fiber}=$ 10 kilometers,
the fiber with the attenuation characteristic of 0.3 dB per kilometer supports an
effective secrecy rate of about 164 bits per second, while the fiber with the
attenuation characteristic of 0.2 dB per kilometer suports this throughput rate out
to a separation distance of 15 kilometers. We also note that at a separation
distance of 25 kilometers, the lower quality fiber supports an effective secrecy rate
of about 57.9 bits per second, while the higher quality fiber supports an effective
secrecy rate of about 103.6 bits per second, corresponding to a gain of about 2.5 dB.

\section{Conclusion}

In this paper we have discussed several features of the practical implementation of
quantum cryptography in real environments
having to do with unconditional secrecy, computational
loads and effective secrecy rates in the presence of perfect and imperfect sources.
As progress in telecommunications and optoelectronics continues to make the insertion
of quantum cryptographic technology into real communications systems a more
realistic prospect it will become increasingly important to uncover further
details and subtleties that determine the optimum possible
performance characteristics.


\section{Acknowledgements}

The authors thank their colleagues in the MITRE Quantum Information Science Group
for valuable discussions. They also wish to thank particular employees of the U.S.
National Security Agency for helpful comments. GG in addition wishes to thank Ja. F.
Providakes for encouragement, and especially D. Lehman and the MITRE Technology Program
for supporting this work.


\end{document}